\documentclass[aps,prb,reprint,superscriptaddress,longbibliography,floatfix]{revtex4-2}

\usepackage{amsmath, amssymb, amsfonts}
\usepackage{bm}

\usepackage{booktabs}
\usepackage{multirow}
\usepackage{graphicx}
\usepackage{hyperref}
\hypersetup{colorlinks=true, linkcolor=blue, citecolor=blue, urlcolor=blue}

\usepackage{siunitx}
\usepackage{xcolor}
\usepackage{microtype}
\usepackage{enumitem}
\usepackage{placeins}
\usepackage{soul}
\renewcommand{\hl}[1]{#1}

\newcommand{\ETE}{\omega}
\newcommand{\wraw}{\omega^{\mathrm{raw}}}
\newcommand{\wcorr}{\omega^{\mathrm{corr}}}
\newcommand{\dEatomall}{\Delta E_{\mathrm{all,atom}}^{\mathrm{AE}}}
\newcommand{\dEatomallPAW}{\Delta E_{\mathrm{all,atom}}^{\mathrm{AE,VASP}}}
\newcommand{\dEatomallUSPP}{\Delta E_{\mathrm{all,atom}}^{\mathrm{AE,CASTEP}}}
\newcommand{\dEatomval}{\Delta E_{\mathrm{val,atom}}}
\newcommand{\dEatomvalUSPP}{\Delta E_{\mathrm{val,atom}}^{\mathrm{CASTEP}}}
\newcommand{\dEcoreatom}{\Delta E_{\mathrm{core,atom}}}
\newcommand{\dEpaw}{\Delta E_{\mathrm{core,atom}}^{\mathrm{VASP}}}

\newcommand{\dEuspp}{\Delta E_{\mathrm{core,atom}}^{\mathrm{CASTEP}}}
\newcommand{\dEvalatomVASP}{\Delta E_{\mathrm{val,atom}}^{\mathrm{VASP}}}
\newcommand{\dEisoVASP}{\Delta E_{\mathrm{iso}}^{\mathrm{VASP}}}
\newcommand{\EtotisoXCH}{E_{\mathrm{iso}}^{\mathrm{FCH,VASP}}}
\newcommand{\EtotisoGS}{E_{\mathrm{iso}}^{\mathrm{GS,VASP}}}
\newcommand{\dEAEref}{\Delta E_{\mathrm{ref,atom}}}
\newcommand{\dEval}{\Delta E_{\mathrm{cell}}}

\newcommand{\ddEval}{\Delta\Delta E_{\mathrm{cell}}}

\newcommand{\eV}{\,\mathrm{eV}}

\makeatletter
\newcommand{\revtexbibstylebeforebibdata}{%
  \expandafter\bibliographystyle@latex\expandafter{\@bibstyle}%
  \gdef\write@bibliographystyle{%
    \@ifxundefined\@bibstyle{}{%
      \bibliographystyle@sw{}{\@bibdataout@rev}%
    }%
    \global\let\write@bibliographystyle\relax
  }%
  \def\selectlanguage##1{}%
}
\makeatother

\begin{document}

\title{All-Electron Single-Atom Reference Correction for Absolute Transition Energies in Fixed-Reference PAW-XCH Calculations}

\author{Yinan Wang}
\email{ynwang20020718@g.ecc.u-tokyo.ac.jp}
\affiliation{Department of Materials Engineering, Graduate School of Engineering, The University of Tokyo, Tokyo, Japan}
\affiliation{Institute of Industrial Science, The University of Tokyo, Tokyo, Japan}

\author{Teruyasu Mizoguchi}
\email{teru@iis.u-tokyo.ac.jp}
\affiliation{Institute of Industrial Science, The University of Tokyo, Tokyo, Japan}

\date{\today}

% ============================================================
\begin{abstract}
\hl{The transition energy of a core-loss spectrum is the energy required to excite a core electron to the lowest unoccupied state, comprising a transferable atomic core-reference contribution and a material-dependent response that gives the chemical shift between different chemical environments. Calculations using density functional theory (DFT) typically place the transition energy below its experimental value. CASTEP can reconstruct the transition energy from its core-electron and valence-electron energy changes. However, the Vienna Ab initio Simulation Package (VASP) uses one fixed PAW dataset for both occupations and retains the ground-state atomic reference when the absorbing-atom core occupation changes.} The resulting transition-energy axis lacks the associated all-electron atomic reference change. \hl{In this study, we evaluate the all-electron shift $\dEatomall$ and the atomic core-reference contribution $\dEcoreatom$ with self-consistent single-atom calculations. The atomic reference contribution missing from the fixed PAW dataset is captured by a residual correction $\dEAEref$, constructed from the single-atom valence reference change and isolated-atom total-energy response obtained with VASP.} This construction gives one functional form for the absorption edges considered. Systematic comparisons with an independent ultrasoft-pseudopotential (USPP) implementation establish the numerical consistency of the all-electron reference over K, L$_1$, and L$_{2,3}$ core holes. \hl{Across most elements, the two implementations reproduce $\dEatomall$, and representative spectra on the corrected axis approach the experimental absolute energy scale.} The direct core-only shift remains dependent on its reference convention. For a fixed element, edge, and core-hole scheme, atomic reference terms cancel from energy differences, so same-element chemical shifts are governed by \hl{the double difference of the supercell total energy,} $\ddEval$. At the Al and Si L$_{2,3}$ edges, the near-edge shape also depends strongly on the PAW dataset and its representation of low-lying $3d$-like unoccupied states.
  \end{abstract}

\maketitle

% ============================================================
\section{Introduction}
\label{sec_intro}
% ============================================================

X-ray absorption near-edge structure (XANES) and electron energy-loss near-edge structure (ELNES) probe local electronic structure, bonding, and coordination around an absorbing element \cite{keast_application_2012,bruley_elnes_1994,muller_connections_1998}. Fingerprint comparison is effective when suitable standards exist, while unfamiliar structures, defects, low-symmetry sites, and multicomponent environments often require first-principles spectra \cite{bruley_elnes_1994,hofer_fundamentals_2016}. Reference libraries may contain several chemically similar candidates, and a visual match alone can leave the microscopic origin of individual features unresolved. A calculated spectrum provides a complementary route from an explicit atomic structure to the observable. It can predict the near-edge line shape and energy position and can relate individual features to unoccupied states, charge redistribution, coordination, and local bonding \cite{tanaka_xanes_2005}.

Energy position is essential when spectra from different phases, oxidation states, or inequivalent sites must be compared on one scale. The absolute onset links a calculated excitation to the experimental photon or energy-loss axis, while a chemical shift provides a local measure within one elemental edge. A theory that supplies both quantities can rank candidate structures before a matching standard is available and can support searchable theoretical spectral databases. These uses require an internally consistent reference because an arbitrary alignment applied separately to each spectrum removes the chemical-shift information of interest.

Theoretical approaches include real-space multiple scattering, as implemented in FEFF \cite{rehr_parameter-free_2010,ankudinov_real-space_1998}, and periodic supercell core-hole calculations \cite{karsai_effects_2018,elsasser_density-functional_2001,jayawardane_cubic_2001}. The latter obtain the core-excited electronic structure within density functional theory. In the \hl{excited} core-hole (XCH) construction, the removed core electron is placed in the conduction manifold, which preserves supercell neutrality \cite{hetenyi_calculation_2004}. Charge neutrality reduces finite-size errors associated with long-range Coulomb interactions between periodic images and makes the method practical for molecules, defects, and solids \cite{fujikata_finite-size_2026}. Plane-wave codes implement related core-hole calculations through projector-augmented-wave (PAW), ultrasoft-pseudopotential (USPP), and other frozen-core frameworks \cite{kresse_efficient_1996,clark_first_2005,gao_core-level_2009,motornyi_simulation_2018,timrov_turboeelscode_2015}. Their atomic energy zeros need not share the same convention even when their periodic valence responses are physically comparable.

Within a supercell approach, the core hole changes both the local potential and the screening charge over the entire simulation cell. The total-energy response contains the chemical environment that is absent from an isolated atomic reference. At the same time, a frozen-core calculation does not vary every core contribution explicitly in the periodic energy functional. A transition-energy construction must combine these two pieces without counting one of them twice. Differences among plane-wave codes largely enter through how the frozen atomic piece and its zero point are stored.

Practical XCH calculations commonly use generalized-gradient-approximation (GGA) functionals such as PBE \cite{langreth_theory_1980,langreth_beyond_1983,perdew_generalized_1996}. Their discrepancies with experiment contain three physically distinct contributions. The absolute reference fixes the energy position of the entire spectrum on an element-specific scale. The same-element chemical shift compares two local environments at a fixed edge and is sensitive to their screened total-energy response. The near-edge shape concerns peak spacing, splitting, and relative intensity \cite{shibata_simulated_2022}. A common rigid displacement can address an energy zero while leaving a chemical-shift error and compressed peak spacing. Conversely, a line-shape transformation does not determine the absolute transition energy. The present work addresses the absolute reference and the same-element chemical shift. GGA spectral compression is associated with self-interaction and electron delocalization and was examined separately in earlier work \cite{wang_systematic_2026}. Absolute positions and chemical shifts remain informative measures of valence, coordination, and bonding \cite{bruley_elnes_1994,muller_connections_1998,tanaka_xanes_2005,keast_application_2012}.

The distinction also determines how experimental spectra are used. Experimental alignment can help display two line shapes on the same plot, but it cannot supply a transferable element-level reference for a new material. The absolute-axis correction must follow from the calculated atomic and supercell energies. Once that axis is defined, spectra of the same element can be compared with one common visual displacement, leaving their calculated relative positions intact.

The core-valence decomposition proposed by Mizoguchi \textit{et al.} provides a total-energy framework for the energy-axis problem \cite{mizoguchi_first-principles_2009}. The transition energy is written as

\hl{Figure~\protect\ref{fig_method_overview}(a) illustrates the reconstruction of the transition energy at the level of core and conduction states. The right-hand side shows the GS atom, with the active core shell fully occupied and the conduction manifold empty. The left-hand side shows the corresponding XCH atom, in which one electron has been removed from the active core shell and placed in the lowest unoccupied conduction orbital, while the supercell remains charge neutral. This pair of levels defines the raw single-particle transition subtracted in Eq.}~\protect\eqref{eq_final_formula}\hl{~and identifies the core and valence contributions entering} $\dEval$ \hl{and the atomic reference terms.}

\begin{equation}
  \ETE = \dEcoreatom + \dEval ,
  \label{eq_energy_conservation}
\end{equation}
$\dEcoreatom$ is the frozen-core atomic contribution and $\dEval$ is the complete self-consistent supercell response to the core hole. The former carries the atomic core reference omitted from the variational valence space, while the latter contains screening and bonding rearrangement in the actual chemical environment. A USPP implementation can retain separate ground-state and core-hole atomic references and form the core contribution from its all-electron and pseudo-valence parts. CASTEP provides an independent USPP reference, and related excited-state pseudopotential constructions are available in Quantum ESPRESSO \cite{motornyi_simulation_2018,timrov_turboeelscode_2015}. These implementations provide a useful atomic comparison because the core-hole pseudopotential explicitly records a changed atomic reference.

\hl{The reliability of this CASTEP atomic reference has been tested directly against experiment. A previous validation of the CASTEP-based correction scheme against gas-phase C K-edge spectra of 44 molecules showed that the calculated first-peak energies followed the experimental trend with a predominantly systematic offset. A common shift of $5.78\eV$ provided a sufficient approximation under a unit-slope assumption, while the chemical shifts among different molecules were quantitatively reproduced} \cite{shibata_simulated_2022}\hl{. These results support the use of CASTEP as a reproducible methodological reference for evaluating the present VASP implementation below, although the CASTEP absolute transition energies retain an approximately $6\eV$ systematic overestimation and are not treated as an error-free experimental standard.}

Unlike CASTEP, VASP use a fixed PAW dataset for both occupations \cite{karsai_effects_2018}. No excited-state PAW dataset is generated when the core occupation changes. The internal one-center core shift leaves the XCH state on the ground-state atomic reference zero point and omits the all-electron atomic reference change associated with core-hole formation.

\hl{Figure~\protect\ref{fig_method_overview}(b) illustrates why retaining the GS pseudopotential for both occupations displaces the computed total energy and requires the residual correction} $\dEAEref$\hl{. Because VASP uses one fixed PAW dataset for the GS and XCH calculations, the internal core reference does not follow the change in core occupation, and the XCH total energy remains tied to the GS atomic zero point. This leaves the raw axis displaced from the true transition energy by an amount not contained in} $\dEval$ \hl{alone. Figure~\protect\ref{fig_method_overview}(b) shows the GS and XCH total-energy levels before and after this reference-zero shift, so that the role of} $\dEAEref$ \hl{in restoring the missing all-electron response is shown directly.}

Figure~\ref{fig_method_overview}(d,e) illustrates the resulting raw absolute positions and same-element chemical-shift errors. We reconstruct the missing contribution with a spherical all-electron single-atom reference and a residual VASP reference correction. The construction is based on atomic reference calculations and remains independent of the visual alignment used in spectral figures. Tests spanning K, L$_1$, and L$_{2,3}$ configurations establish numerical consistency and identify the applicable edge classes. The corrected absolute axis is then examined for C, N, O, and F K edges, while Al and Si L$_{2,3}$ spectra provide a limiting test of PAW-dataset sensitivity.

\section{Theory and Methods}
\label{sec_theory}
% ============================================================

\subsection{Total-Energy Decomposition of the Transition Energy}
\label{sec_decomp}

All energy differences use the same sign convention, the core-hole state minus the ground state (GS),
\begin{equation}
  \Delta E = E^{\mathrm{XCH}} - E^{\mathrm{GS}} .
  \label{eq_sign_convention}
\end{equation}
For the absorption site under consideration, $\dEval$ denotes the complete self-consistent supercell contribution under this convention.
It contains the screening, bonding rearrangement, and chemical-environment response of the actual supercell under the fixed VASP PAW reference. The difference is evaluated separately for every inequivalent absorption site because the relaxed XCH charge density and total energy depend on the local environment. It includes all self-consistent valence contributions in the periodic molecule or solid and is distinct from a single-atom pseudo-valence reference. The atomic term $\dEcoreatom$ accounts for the frozen-core contribution and depends on the reference convention, as defined in Sec.~\ref{sec_atom_core}. Explicit excited pseudopotentials carry separate GS and XCH atomic references \cite{mizoguchi_first-principles_2009}, which allows the atomic contribution to be reconstructed from reference total energies. The independent USPP calculations used below were performed with CASTEP \cite{clark_first_2005}.

Equation~\eqref{eq_energy_conservation} separates a transferable atomic edge reference from a material-specific screened response. The separation is useful only after a reference convention has been fixed for the atomic part. Within one convention, $\dEval$ can be compared between sites because the same PAW dataset and core-hole prescription are used. For two environments of the same elemental edge, the atomic contribution is common and the double difference of the supercell energies gives the chemical shift. For different elements or edges, the full atomic contribution must be retained to establish an absolute transition-energy axis.

\begin{figure*}[!p]
  \centering
  \includegraphics[width=0.88\textwidth,height=0.86\textheight,keepaspectratio]{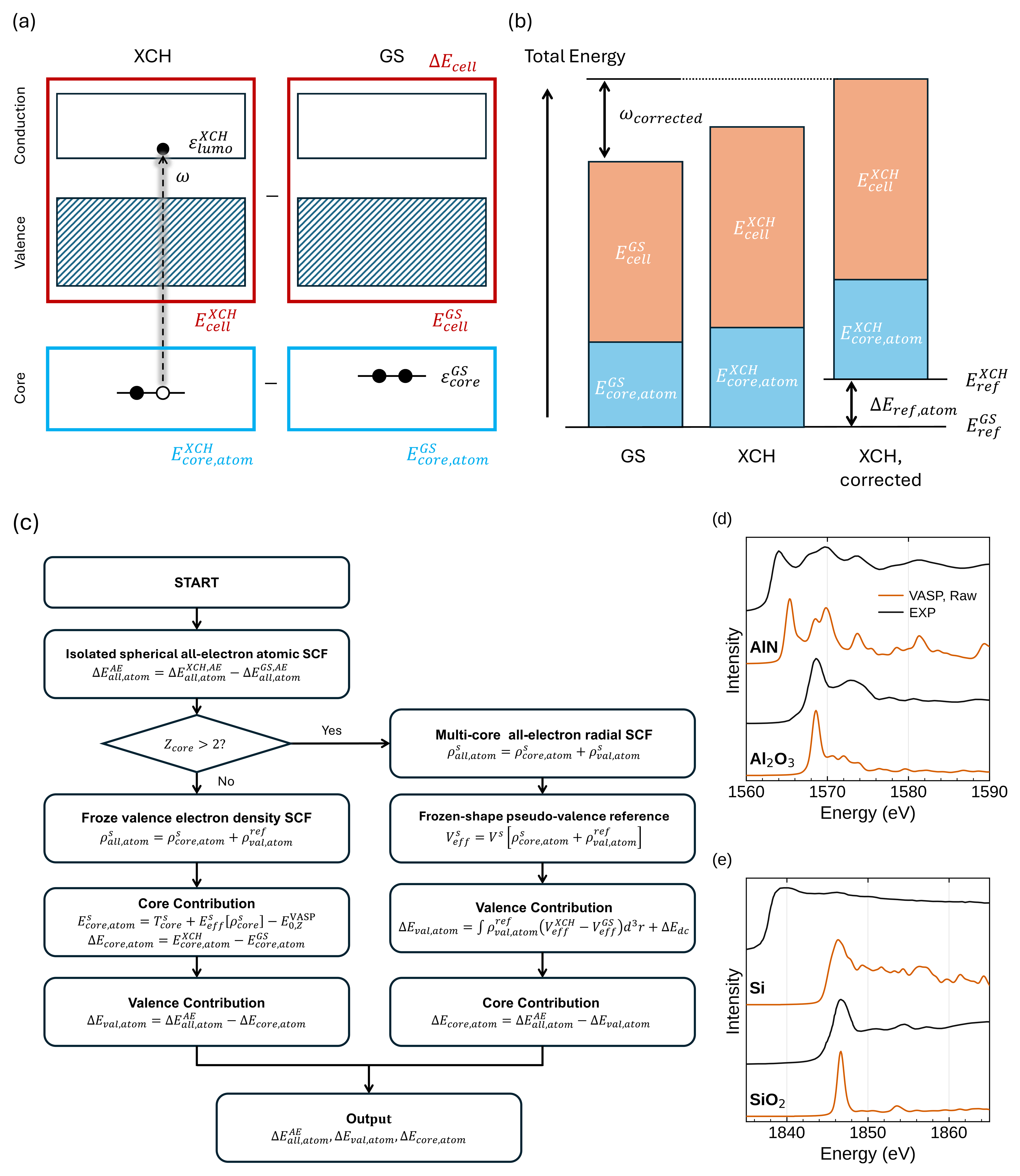}
  \caption{Overview of the transition-energy construction. (a)~Core and valence levels entering Eq.~\eqref{eq_energy_conservation}, including $\dEval$ and the raw single-particle transition. (b)~GS and XCH total-energy levels before and after the reference-zero shift $\dEAEref$. (c)~Single-atom workflow for $\dEatomall$, $\dEpaw$, and $\dEAEref$. (d) and (e)~Raw VASP-XCH spectra for the Al K edge of w-AlN \cite{balasubramanian_characterization_2006} and $\alpha$-Al$_2$O$_3$ \cite{murao_thermodynamic_2018}, and the Si K edge of bulk Si and $\alpha$-SiO$_2$ \cite{wu_facile_2017}. The computed oxide spectrum in each pair is aligned to experiment for visualization and the same energy shift is applied to the paired spectrum.}
  \label{fig_method_overview}
\end{figure*}

\hl{Figure~\protect\ref{fig_method_overview}(c) shows the computational workflow added to the standard VASP procedure for the single-atom approximation. Starting from the same PAW dataset and core-hole occupation used in the periodic calculation, this workflow evaluates the spherical all-electron atom to obtain} $\dEatomall$\hl{, evaluates the fixed-dataset one-center core reference to obtain} $\dEpaw$\hl{, and combines the isolated-atom VASP total energies to obtain} $\dEAEref$ \hl{through Eq.}~\protect\eqref{eq_ae_ref_final}\hl{. These atomic calculations are independent of the periodic supercell and are performed once per element, edge, and core-hole scheme.}

\subsection{All-Electron Single-Atom Reference and VASP Reference Correction}
\label{sec_atom_all}
\label{sec_atom_core}

Quantities with an ``atom'' or ``iso'' subscript refer to single-atom or isolated-atom references and are distinct from the periodic $\dEval$. The all-electron shift $\dEatomall$ follows Eq.~\eqref{eq_sign_convention} and is obtained from self-consistent spherical calculations of the absorbing atom in its GS and core-hole occupations. The occupations are constructed from the PAW atomic configuration. The target core occupation is reduced by the prescribed hole charge without a compensating valence electron, so this reference describes an ionized core-hole atom. The nuclear charge remains the elemental value $Z$ and is passed independently of the occupation-derived electron number. Coupling these inputs would instead change the element represented by the radial problem and would displace its total energy by hundreds of eV. Active and spectator core shells and occupied valence shells relax in the same atomic potential, which includes spectator-core and valence relaxation in $\dEatomall$. Appendix~\ref{app_S1} gives the radial construction and total-energy expression.

Both atomic states use the same exchange-correlation functional, relativistic level, radial boundary convention, and occupation rule. The GS calculation begins from the neutral atomic configuration. The core-hole calculation changes only the prescribed active-shell occupation and then relaxes all occupied radial orbitals self-consistently. This construction keeps the all-electron reference tied to a single element while allowing the core, spectator, and valence densities to respond to the hole. The resulting $\dEatomall$ is a complete atomic total-energy change and does not require a separate choice of how to distribute core-valence interaction terms.

The pseudo-valence single-atom reference shift is $\dEatomval$. We use $\dEvalatomVASP$ and $\dEatomvalUSPP$ when the implementation matters, with the generic symbol referring to the VASP branch. The USPP partition is
\begin{equation}
  \dEuspp = \dEatomallUSPP - \dEatomvalUSPP ,
  \label{eq_uspp_partition}
\end{equation}
where $\dEatomallUSPP$ has the all-electron definition above. The partition subtracts the pseudo-valence atomic reference associated with the explicit core-hole pseudopotential. Coulomb and exchange-correlation energies contain core-valence cross terms, and the Hartree energy also depends on the combined core and valence density. Assigning these cross terms to one part of a core-valence decomposition requires a reference convention. A core-only shift can consequently vary across implementations even when the complete all-electron atomic total-energy change agrees closely. This distinction is central to the comparisons in Sec.~\ref{sec_atomref_validation}.

The convention dependence is expected from the structure of the density functional. A core density and a valence density each generate electrostatic and exchange-correlation contributions that are nonlinear in their sum. Moving a cross term from the core part to the valence part leaves the complete atomic energy unchanged while changing the reported core-only energy. Comparisons of $\dEuspp$ with a VASP one-center quantity test the reference partition. Comparisons of $\dEatomall$ test the underlying all-electron atomic response.

The direct VASP one-center core-reference shift is $\dEpaw$, where the one-center functional is specified in Appendix~\ref{app_S1_paw_transform}. The functional uses the spherically averaged PAW valence reference density as a fixed background while the core shells are solved for the selected occupation. It yields a direct change of the VASP one-center core reference. This construction differs from the USPP partition of Eq.~\eqref{eq_uspp_partition}, which subtracts an independently defined pseudo-valence reference.

The PAW dataset also contains an element-specific reference constant. Because the same dataset is used for the two occupations, this constant cancels in $\dEpaw$. The fixed valence background still affects the effective potential, core eigenvalues, and kinetic contribution. The remaining Hartree, electron-nucleus, and exchange-correlation terms are evaluated from the core density according to the one-center functional. These choices reproduce the internal VASP reference convention required for a consistent correction.

The fixed PAW dataset leaves an additional all-electron reference change outside this direct one-center shift. Its construction uses an isolated-atom full core-hole (FCH) occupation without valence compensation. In FCH, the removed core electron is not placed in the lowest unoccupied molecular orbital (LUMO) of a molecule or the conduction-band minimum of a solid. The charge-neutral XCH construction instead places this electron in the conduction manifold. The corresponding VASP total-energy shift is
\begin{equation}
  \dEisoVASP
  = \EtotisoXCH
  - \EtotisoGS ,
  \label{eq_iso_atom_def}
\end{equation}
where both states use the same converged vacuum supercell, exchange-correlation functional, and VASP dataset. The FCH choice matches the electron number in the all-electron ionized atom, so the pseudopotential and all-electron atoms represent the same occupation change. The associated VASP single-atom valence reference change is $\dEvalatomVASP$. It follows the same FCH construction and the internal VASP pseudo-valence convention. The residual correction is
\begin{equation}
  \dEAEref
  = \dEvalatomVASP
  - \dEisoVASP ,
  \label{eq_ae_ref_final}
\end{equation}
Equation~\eqref{eq_ae_ref_final} combines the VASP valence reference change and total-energy response of the same isolated atom. The two terms share the element, PAW dataset, functional, and core-hole occupation, which keeps the subtraction within one VASP reference convention. One functional form applies to every edge considered. For a fixed element, absorption edge, and core-hole scheme, $\dEAEref$ is an element-level constant obtained entirely from atomic reference calculations. It contains no fitted experimental onset. The constant shifts the absolute axis and cancels between chemical environments of the same element and edge. The light-element and multi-core numerical branches determine how the underlying atomic problem is solved and are introduced in Sec.~\ref{sec_branch_logic}.

The direct term $\dEpaw$ and residual term $\dEAEref$ have complementary roles. The former follows the one-center core functional stored by the PAW description. The latter restores the remaining atomic zero-point response associated with the changed occupation. Their sum supplies the atomic contribution on the final axis. Recomputing both terms under a changed dataset or occupation convention keeps the construction internally consistent.

\subsection{Corrected Transition-Energy Axis}
\label{sec_final_formula}

Let $\wraw$ denote the raw XCH spectral axis for one absorption site. Each feature is initially positioned by a Kohn-Sham eigenvalue difference from the ground-state core level to an unoccupied XCH state. The reference transition connects $\varepsilon_{\mathrm{core}}^{\mathrm{GS}}$ to the lowest unoccupied XCH state $\varepsilon_{\mathrm{lumo}}^{\mathrm{XCH}}$. The latter is the conduction-band minimum in a periodic solid or the corresponding LUMO-like state in a molecular supercell. This eigenvalue difference supplies a convenient internal origin for the raw spectrum, while the total-energy decomposition supplies the transition-energy reference. \hl{The raw VASP-computed spectrum carries an intrinsic displacement built from this internal single-particle origin, and this displacement must be removed to align the calculated onset with the transition-energy reference. The required correction is $\varepsilon_{\mathrm{lumo}}^{\mathrm{XCH}}-\varepsilon_{\mathrm{core}}^{\mathrm{GS}}$, obtained directly from the ground-state and core-hole Kohn-Sham eigenvalues of the same calculation.} Replacing the former with the latter gives
\begin{align}
  \wcorr
  &= \wraw
  - \left(\varepsilon_{\mathrm{lumo}}^{\mathrm{XCH}}
  - \varepsilon_{\mathrm{core}}^{\mathrm{GS}}\right) \nonumber\\
  &\quad + \dEval + \dEpaw + \dEAEref .
  \label{eq_final_formula}
\end{align}
Equation~\eqref{eq_final_formula} applies a rigid energy shift to the complete site-resolved spectrum. Every peak from the same absorbing site receives the same displacement. At the lowest unoccupied reference state, the corrected onset equals the transition energy $\ETE$ of Eq.~\eqref{eq_energy_conservation}. The construction changes the absolute position and site-to-site placement of spectra through total energies. It leaves oscillator strengths, broadening, and relative peak spacing unchanged, so GGA spectral compression remains a separate problem.

For a molecule with several inequivalent atoms, Eq.~\eqref{eq_final_formula} is evaluated for each absorbing site using its own core eigenvalue, lowest unoccupied reference, and $\dEval$. The element-level atomic terms are reused when the element, edge, and core-hole scheme are unchanged. Summing the shifted site spectra then preserves the site-resolved matrix elements while placing their onsets according to the corresponding total-energy differences. The same procedure applies to cross-material comparisons after separate periodic calculations have been completed.

\subsection{Evaluation of the Atomic Reference for the Two Edge Classes}
\label{sec_branch_logic}

The all-electron reference is evaluated in two branches distinguished by $Z_{\mathrm{core}}$, the number of core electrons below the active hole. For $Z_{\mathrm{core}}>2$, as at the Al and Si K, L$_1$, and L$_{2,3}$ edges, spectator core shells accompany the active hole. The active core shell, spectator shells, and occupied valence shells relax self-consistently in one spherical all-electron potential. Their coupled response is included in $\dEatomall$. For the C, N, O, and F K edges, the active $1s$ shell changes from $1s^2$ to $1s^1$. Its strong localization and the absence of deeper spectator shells give a stable radial reference. The direct VASP comparison quantity in both branches is $\dEpaw$, while the USPP comparison uses Eq.~\eqref{eq_uspp_partition}. Equation~\eqref{eq_ae_ref_final} retains the same form after either branch has supplied its atomic quantities.

Treating the residual reference as an element-level term is a controlled approximation. It assumes that the missing atomic zero-point change is fixed by the element, edge, and prescribed core-hole scheme, while chemical-environment dependence remains in $\dEval$. The approximation is best conditioned for the light-element K edges because their compact $1s$ orbital is insensitive to the outer boundary of the radial grid. Shallow bound valence states can make the eigenvalue search unstable, and dataset-dependent pseudo-valence channel ordering can prevent a valid initialization. These limitations are listed in Sec.~\ref{sec_limits}. Appendix~\ref{app_S1} gives the detailed solver, convergence, and integration procedures.

\subsection{Core-Hole Construction for the L\texorpdfstring{$_{2,3}$}{2,3} Edge}
\label{sec_l23_method}

The Al and Si L$_{2,3}$ edges involve a $2p$ core hole with angular-momentum character and spin-orbit-split L$_2$ and L$_3$ initial states. Experiment associates these components with the $2p_{1/2}$ and $2p_{3/2}$ levels. \hl{This spin-orbit splitting is only about $0.4\eV$ for Al and $0.6\eV$ for Si} \cite{shinotsuka_bayesian_2025,preobrajenski_honeycomb_2021, kimoto_study_2003}\hl{, comparable to the spectral broadening applied throughout this work, so the L$_2$ and L$_3$ near-edge responses overlap substantially in the computed and experimental spectra.} The charge-neutral supercell calculation removes one electron from the $2p$ shell, promotes it to the conduction manifold, and provides a combined $2p$-hole spectral response and $\dEval$. The single-atom calculation supplies a scalar, spherically averaged reference for that merged response. The present implementation does not construct separate $2p_{1/2}$ and $2p_{3/2}$ all-electron final-state references. A spin-orbit-resolved absolute axis would also require a non-spherical core-hole treatment and consistent relativistic reference. Al and Si L$_{2,3}$ edges are consequently excluded from the primary absolute-energy validation and are used to examine PAW-dataset sensitivity in Sec.~\ref{sec_l23_results}.

\section{Results}
\label{sec_results}
% ============================================================

\subsection{Single-Atom All-Electron Reference Validation}
\label{sec_atomref_validation}

\begin{figure}[!t]
  \centering
  \includegraphics[width=\columnwidth]{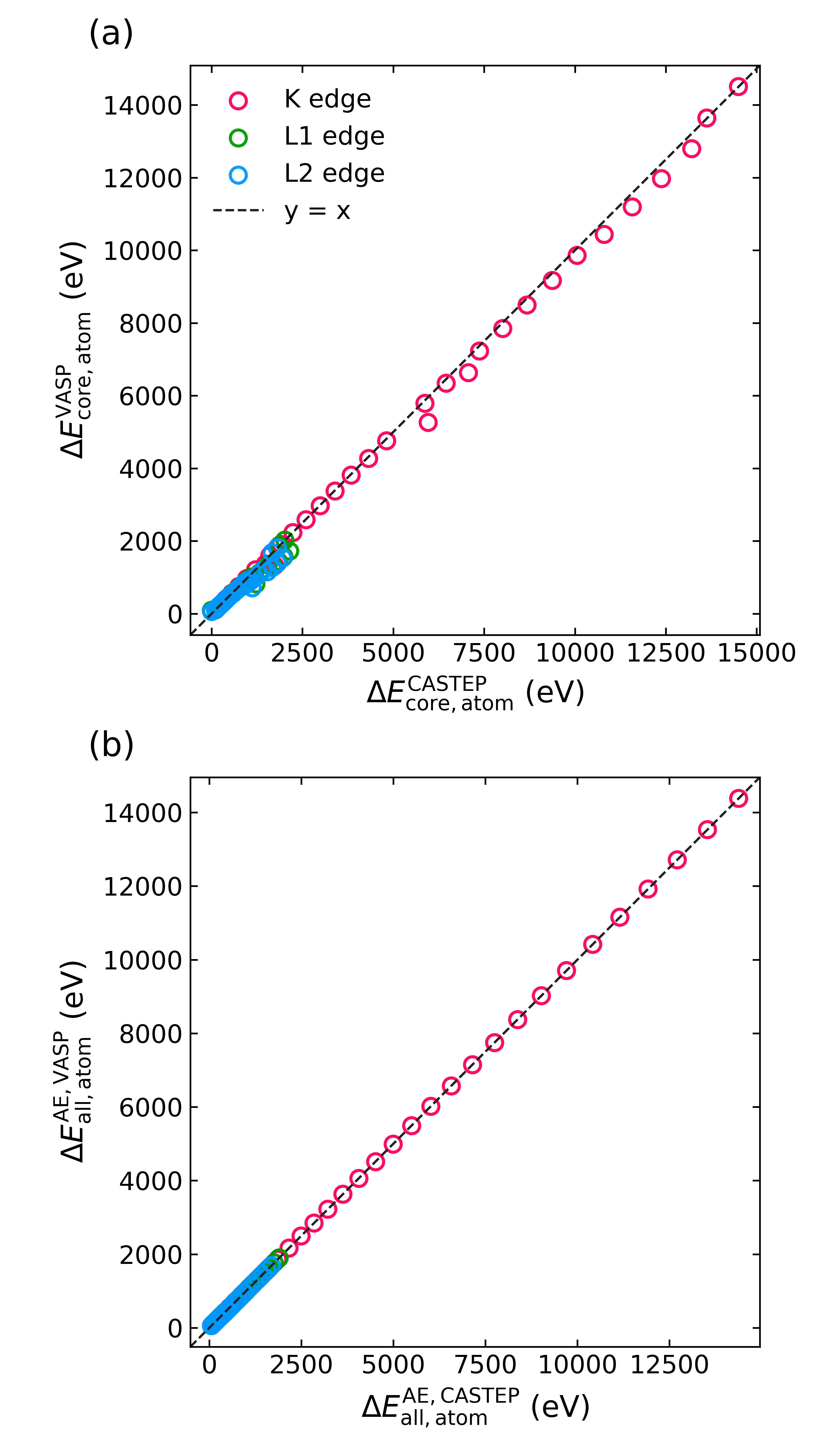}
  \caption{Single-atom reference comparison for K, L$_1$, and L$_{2,3}$ core-hole configurations from the second through the fourth periods. (a)~Direct VASP one-center shift $\dEpaw$ against the USPP partition $\dEuspp$. (b)~All-electron single-atom shift $\dEatomall$ in the two implementations. The dashed line denotes $y=x$. Li K, Be K, B K, Na L$_1$, and Na L$_{2,3}$ are excluded as described in Sec.~\ref{sec_limits}. The plotted L$_2$ series represents the $2p$ core-hole shell.}
  \label{fig_parity}
\end{figure}

Figure~\ref{fig_parity} summarizes single-atom tests for K, L$_1$, and L$_{2,3}$ core-hole configurations from the second through the fourth periods. This range samples active holes with different principal and angular-momentum quantum numbers and atoms with progressively more spectator core shells. The plotted L$_2$ series represents the unresolved $2p$ core-hole shell. Li K, Be K, B K, Na L$_1$, and Na L$_{2,3}$ are excluded because the auxiliary construction does not return a valid reference, as discussed in Sec.~\ref{sec_limits}. These exclusions do not originate from the periodic XCH calculations or the USPP reference data. The remaining set provides a broad numerical test of the radial atomic procedure before any comparison with experimental transition energies.

Panel (b) compares $\dEatomall$, which is the total-energy change of the same spherical core-hole atom in both implementations. The nuclear charge, active-shell occupation change, and ionized electron number have the same physical definition on the two axes. Agreement across three periods and three core-hole types establishes the numerical consistency of the all-electron atomic solver and shows that spectator-core relaxation is reproduced over a wide range of atomic configurations. This is a solver-level test of the atomic reference. Absolute transition energies additionally contain the material-dependent $\dEval$ and are evaluated below through edge-specific comparisons with experiment.

Panel (a) compares quantities with different reference conventions. The USPP axis is the partition $\dEuspp$ from Eq.~\eqref{eq_uspp_partition}, while the PAW axis is the direct one-center shift $\dEpaw$. The former subtracts a pseudo-valence atomic total-energy change. The latter evaluates a core functional in the fixed spherical PAW valence background. Departures from $y=x$ reflect the different core-valence partitions. Their growth for heavier elements is consistent with a larger role of spectator shells and dataset-dependent pseudo-valence definitions. The corrected C, N, O, and F K-edge axes are tested against representative experimental spectra in Sec.~\ref{sec_corrected_abs}.

\subsection{Atomic Reference Validation and Residual Correction at the K Edges}
\label{sec_core_anchor}
\label{sec_deref_results}

Table~\ref{tab_core_anchor} compares the all-electron shift $\dEatomall$ from the PAW and independent USPP implementations. Across the listed edges, the two implementations differ by at most about $0.02\eV$ for C, N, O, and F and about $0.04\eV$ for Si, P, S, and Cl. This agreement establishes that the complete all-electron single-atom response is numerically reproducible across the tested K edges.

\begin{table}[t]
\caption{All-electron single-atom shift $\dEatomall$ from the PAW and independent USPP implementations for the listed K edges. The difference is PAW minus USPP and was evaluated before rounding the listed values.}
\label{tab_core_anchor}
\begin{ruledtabular}
\begin{tabular}{lrrr}
Edge &
$\dEatomallPAW$ &
$\dEatomallUSPP$ &
Difference \\
 & (eV) & (eV) & (eV) \\
\hline
C\,K  & 304.39  & 304.38  & $+0.01$ \\
N\,K  & 420.28  & 420.27  & $+0.01$ \\
O\,K  & 554.84  & 554.83  & $+0.01$ \\
F\,K  & 708.11  & 708.10  & $+0.02$ \\
Si\,K & 1861.68 & 1861.65 & $+0.03$ \\
P\,K  & 2166.40 & 2166.37 & $+0.04$ \\
S\,K  & 2494.64 & 2494.60 & $+0.04$ \\
Cl\,K & 2846.48 & 2846.44 & $+0.04$ \\
\end{tabular}
\end{ruledtabular}
\end{table}

The C, N, O, and F cases involve a compact $1s$ orbital and no deeper spectator core. Their core-hole response is dominated by the radial density close to the nucleus, and the $1s^2\rightarrow1s^1$ occupation change is represented directly in both atomic solvers. This combination makes the spherical reference especially stable and explains the approximately $0.02\eV$ consistency. These four edges form the primary set for the absolute-energy comparisons below. The Si, P, S, and Cl values provide an additional multi-core check in which the spectator shells relax with the active hole.

The small differences also provide a scale for interpreting later reference terms. They are several orders of magnitude smaller than the atomic shifts themselves and much smaller than the spectral displacements required to place raw PAW-XCH results on an absolute axis. The agreement isolates the missing zero-point problem from numerical uncertainty in the spherical all-electron solver. It does not constrain the material-dependent supercell response, which must be tested through actual spectra.

Table~\ref{tab_core_anchor_boundary} compares the direct VASP one-center shift $\dEpaw$ with the USPP partition $\dEuspp$. The same VASP construction is used for every row, while the USPP quantity subtracts a different pseudo-valence reference. For the light K edges the difference ranges from $-0.61$ to $+3.56\eV$. It changes sign beyond Si and reaches $-19.89\eV$ at Cl, consistent with the broader trend in Fig.~\ref{fig_parity}. The complete all-electron shifts in Table~\ref{tab_core_anchor} remain closely matched. The spread measures the reference convention used to isolate a core-only quantity and does not indicate a disagreement in the complete atomic total-energy response.

\begin{table}[t]
\caption{Direct VASP one-center shift $\dEpaw$ and USPP core-reference partition $\dEuspp$ for the listed K edges. The difference is VASP minus USPP and was evaluated before rounding the listed values.}
\label{tab_core_anchor_boundary}
\begin{ruledtabular}
\begin{tabular}{lrrr}
Edge &
$\dEpaw$ &
$\dEuspp$ &
Difference \\
 & (eV) & (eV) & (eV) \\
\hline
C\,K  & 399.47  & 397.66  & $+1.81$  \\
N\,K  & 560.71  & 557.15  & $+3.56$  \\
O\,K  & 749.24  & 747.51  & $+1.73$  \\
F\,K  & 965.10  & 965.71  & $-0.61$  \\
Si\,K & 1904.06 & 1902.31 & $+1.75$  \\
P\,K  & 2231.63 & 2238.18 & $-6.55$  \\
S\,K  & 2586.28 & 2600.28 & $-14.01$ \\
Cl\,K & 2969.54 & 2989.43 & $-19.89$ \\
\end{tabular}
\end{ruledtabular}
\end{table}

Table~\ref{tab_deref_main} gives the residual correction $\dEAEref$ from Eq.~\eqref{eq_ae_ref_final}. Its magnitude increases from $53.44\eV$ for C to $133.87\eV$ for Cl across the listed K edges. The negative sign follows the common XCH-minus-GS convention and the ordering of terms in Eq.~\eqref{eq_ae_ref_final}. This residual supplies the element-level zero-point change left outside the direct VASP one-center shift. It modifies the absolute transition-energy axis after the internal single-particle reference has been replaced. For a fixed element and K edge, the same value applies across chemical environments and cancels from the differences in Sec.~\ref{sec_chemical_shift}.

\begin{table}[t]
\caption{Residual atomic reference correction $\dEAEref$ from Eq.~\eqref{eq_ae_ref_final} for the listed K edges.}
\label{tab_deref_main}
\begin{ruledtabular}
\begin{tabular}{lr}
Edge & $\dEAEref$ (eV) \\
\hline
C\,K  & $-53.44$ \\
N\,K  & $-63.21$ \\
O\,K  & $-73.11$ \\
F\,K  & $-84.03$ \\
Si\,K & $-114.11$ \\
P\,K  & $-120.48$ \\
S\,K  & $-126.65$ \\
Cl\,K & $-133.87$ \\
\end{tabular}
\end{ruledtabular}
\end{table}

\label{sec_decore_discussion}
Together, Tables~\ref{tab_core_anchor} to \ref{tab_deref_main} separate the reproducible all-electron atomic response from convention dependence of a core-only partition. The first comparison verifies that two solvers recover the same complete atomic energy change. The second shows how a chosen pseudo-valence reference redistributes energy between core and valence parts. Equation~\eqref{eq_ae_ref_final} then converts the fixed VASP reference into the residual term required by the absolute axis using atomic calculations alone. This sequence avoids treating the convention-dependent core-only difference as a transferable observable. The construction preserves the same-element chemical shift because its element-level term cancels in that difference.

The three tables serve different purposes. Table~\ref{tab_core_anchor} tests a common all-electron definition. Table~\ref{tab_core_anchor_boundary} quantifies how two frozen-core implementations partition that atomic energy. Table~\ref{tab_deref_main} supplies the numerical residual used with the direct VASP term. Keeping these roles separate prevents the close agreement in the first table from being interpreted as a direct test of the complete solid-state transition energy.

\subsection{Corrected Absolute Transition Energies}
\label{sec_corrected_abs}

\begin{figure}[!t]
  \centering
  \includegraphics[width=\columnwidth,height=0.85\textheight,keepaspectratio]{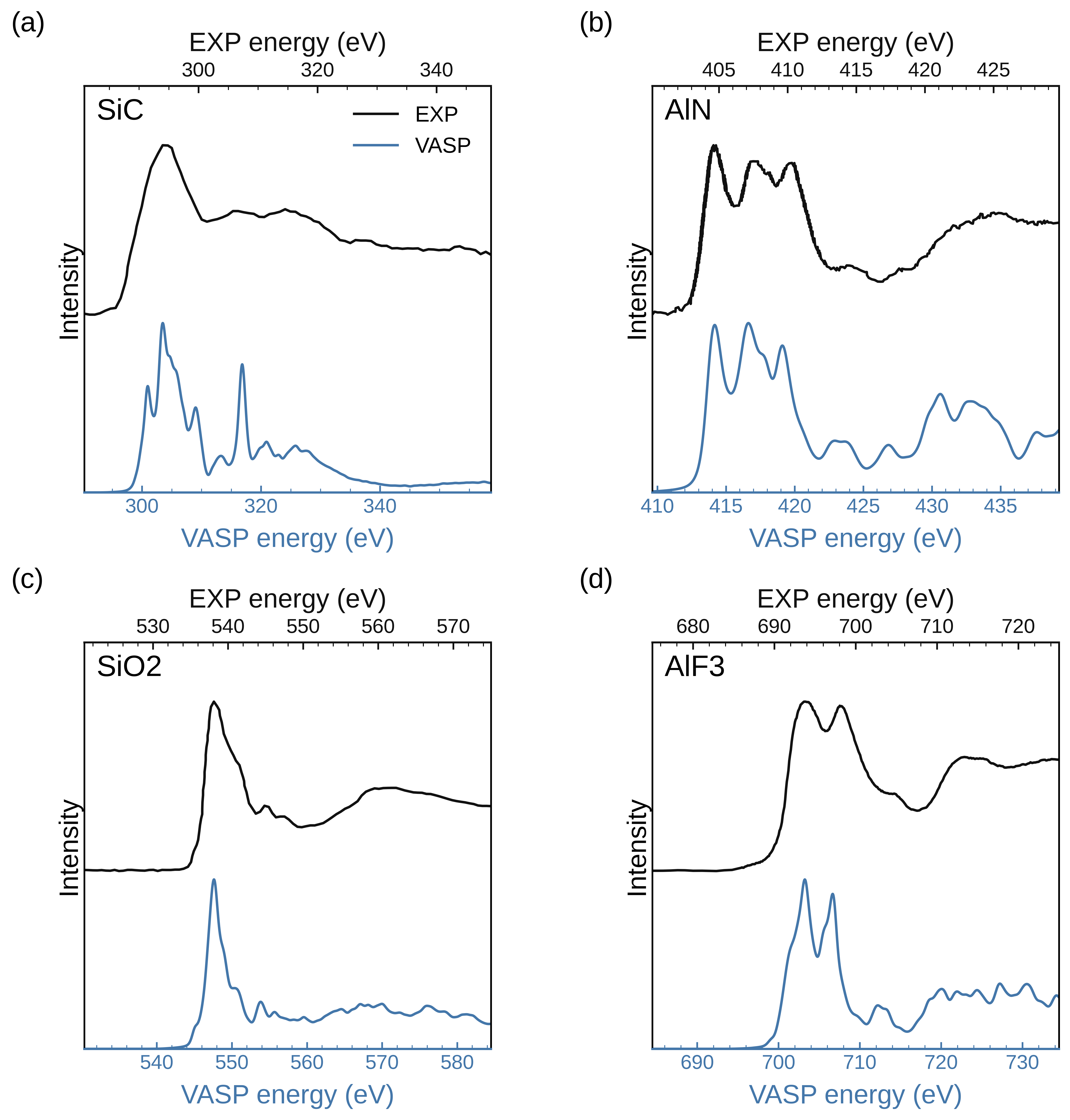}
  \caption{Experimental spectra in black and corrected PAW-XCH spectra in blue. (a)~C K-edge ELNES of 3C-SiC \cite{fu_microstructure_2008}. (b)~N K-edge ELNES of w-AlN \cite{mizoguchi_theoretical_2003}. (c)~O K-edge ELNES of $\alpha$-SiO$_2$ \cite{garvie_can_2010}. (d)~F K-edge XANES of $\alpha$-AlF$_3$ \cite{ishii_integration_2023}. \hl{Each panel is plotted against a lower axis giving the corrected transition energy of Eq.}~\protect\eqref{eq_final_formula}\hl{~and an upper axis giving the experimental transition energy, offset so that the first near-edge peak of the two spectra aligns visually for comparison. This visual alignment is separate from the construction of Eq.}~\protect\eqref{eq_final_formula}.}
  \label{fig_corrected_spectra}
\end{figure}

Figure~\ref{fig_corrected_spectra} compares representative C, N, O, and F K-edge spectra after applying Eq.~\eqref{eq_final_formula}. The four elements span the light K-edge range for which Table~\ref{tab_core_anchor} shows the closest atomic-reference consistency. In each case, replacement of the single-particle origin and addition of $\dEval$, $\dEpaw$, and $\dEAEref$ move the calculated edge toward the experimental absolute energy scale. The agreement is most relevant at the onset and first near-edge features because these locate the corrected transition energy. \hl{Each panel reports this comparison with a lower axis giving the corrected transition energy of Eq.}~\protect\eqref{eq_final_formula}\hl{~and an upper axis giving the experimental transition energy, with the two axes offset so that the first near-edge peak of the calculated and experimental spectra align visually. The resulting offset between the corrected and experimental peak position is $-1.78\eV$ for 3C-SiC, $-3.71\eV$ for w-AlN, $-8.39\eV$ for $\alpha$-SiO$_2$, and $-9.49\eV$ for $\alpha$-AlF$_3$, corresponding to $-0.61\%$, $-0.92\%$, $-1.56\%$, and $-1.37\%$ of the respective absolute experimental transition energy. Given the several-eV width of the first near-edge feature under the applied broadening, a relative offset at this level is small and supports the approach of the corrected axis to the experimental absolute energy scale.} Residual differences remain in high-energy peak spacing, relative intensity, and broadening. Several computed peak groups occupy a narrower energy interval, and experimental lifetime and instrumental widths are only approximated by the common broadening model. These spectral-shape errors include GGA compression of the unoccupied-state energy span and differences between computed and experimental broadening, which a rigid energy-axis correction does not modify.

The same construction is used for all four panels, so no edge-specific experimental offset enters Eq.~\eqref{eq_final_formula}. The comparison tests whether the atomic and supercell terms together recover a physically meaningful absolute position across different elements and bonding environments. The remaining line-shape differences set the appropriate accuracy boundary for interpreting individual high-energy peaks.

The tested systems also separate the atomic reference from the material response. Each panel uses a different bonding environment and a different value of $\dEval$, while the atomic correction is determined only by its elemental K edge. Agreement across this set tests the combined construction under several molecular and solid-state screening environments.

\subsection{Chemical Shift for Intramolecular Sites and Cross-Material Comparison}
\label{sec_chemical_shift}

For systems $A$ and $B$ with the same element, absorption edge, and core-hole scheme, the atomic reference terms in Eq.~\eqref{eq_final_formula} are identical. Their chemical shift is
\begin{equation}
  \Delta\omega_{A-B}
  = \dEval^{A}-\dEval^{B}
  = \ddEval .
  \label{eq_chemical_shift}
\end{equation}
All comparisons in this section use this relation.

The double difference retains the full screened response of each supercell. It includes charge redistribution around the core hole, changes in bonding, and finite-cell contributions treated consistently in the two calculations. The relation requires identical elemental, edge, and core-hole conventions for the compared spectra. When these conditions hold, no separate atomic term is needed to rank their relative onsets.

\begin{figure}[!t]
  \centering
  \includegraphics[width=\columnwidth,height=0.65\textheight,keepaspectratio]{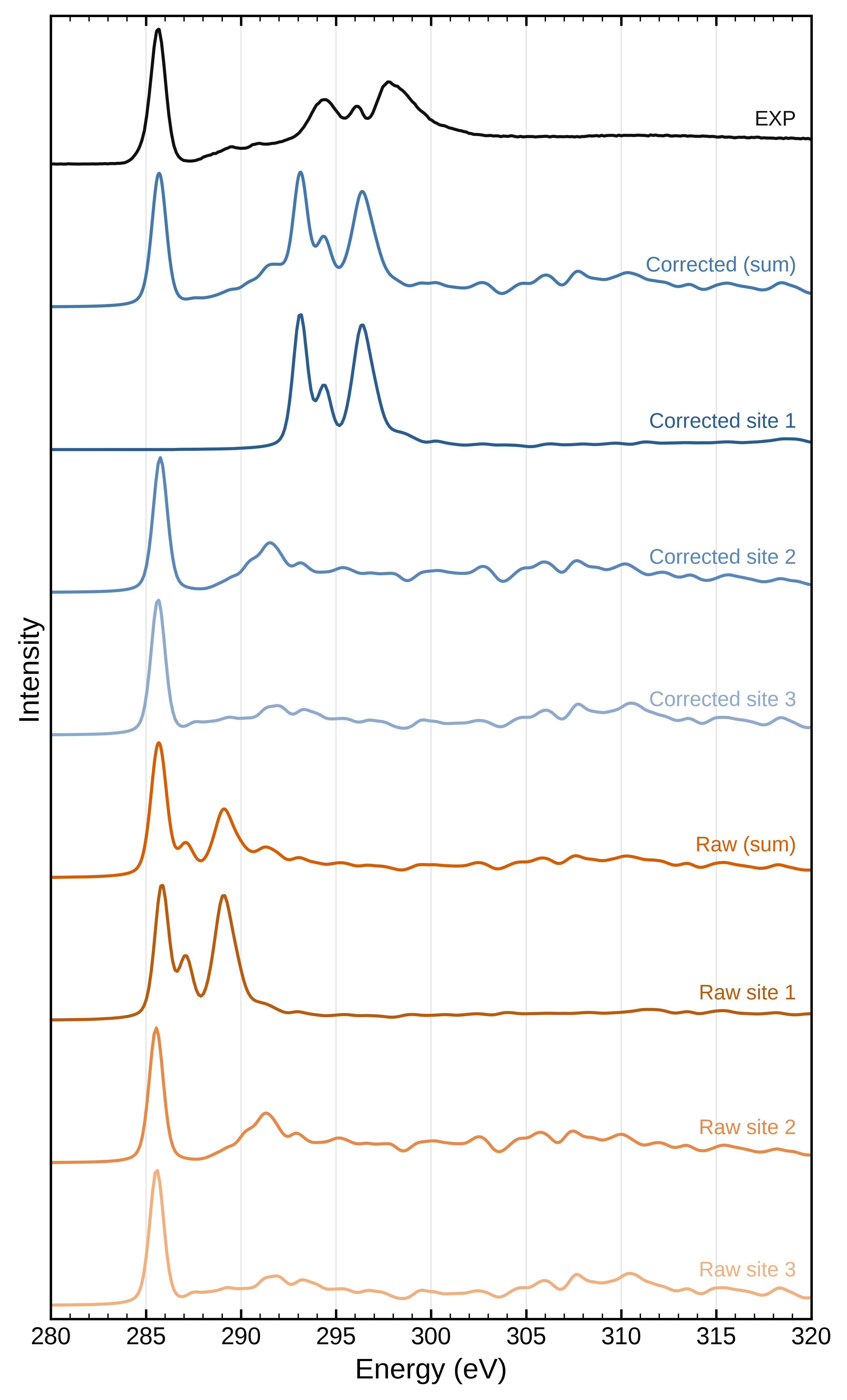}
  \caption{C K-edge chemical shift in 1,1,1-trifluoropropyne. From top to bottom, the curves are the experimental ELNES spectrum \cite{hitchcock_inner_2000}, the corrected sum, corrected sites 1 to 3, the raw sum, and raw sites 1 to 3. Within each group the summed spectrum is aligned to the first experimental peak and the same energy shift is applied to its site-resolved spectra. The experimental alignment is used only for visualization and is separate from the construction of Eq.~\eqref{eq_final_formula}.}
  \label{fig_chemical_shift}
\end{figure}

1,1,1-Trifluoropropyne (CF$_3$-C$\equiv$CH) contains a CF$_3$ carbon, an adjacent alkyne carbon, and a terminal alkyne carbon. These sites sample a strongly fluorinated carbon and two chemically similar alkyne environments within one molecule. The raw spectra retain site-dependent single-particle level differences, while the corrected spectra use $\ddEval$ to set their relative energy positions. Each site keeps its independently calculated near-edge matrix elements and line shape. Only the site energy origin changes when Eq.~\eqref{eq_final_formula} is applied.
The site-resolved $\dEval$ values are $-45.80$, $-53.16$, and $-53.27\eV$ for the CF$_3$-group carbon (site 1), the alkyne carbon bonded to CF$_3$ (site 2), and the terminal alkyne carbon (site 3).
Using the unrounded total energies and referencing the CF$_3$-group carbon, the double differences are $\ddEval = -7.36\eV$ for site 2 and $-7.48\eV$ for site 3, so the two alkyne carbons lie close together, about $7.40\eV$ below the CF$_3$ carbon, and are separated from each other by only $0.11\eV$.
This ordering produces a clearer separation of site-resolved features near 293--298\,eV in Fig.~\ref{fig_chemical_shift}. The two alkyne sites remain nearly coincident on the corrected axis, while their group is separated from the CF$_3$ carbon by about $7.40\eV$. The summed spectrum reflects a chemically interpretable site ordering in place of the smaller raw single-particle separation.

The close values for sites 2 and 3 are consistent with their shared alkyne framework, while the fluorinated site experiences a substantially different local potential and screened response. The calculation resolves this contrast through total energies without altering the distinct site spectra. The comparison also shows that a molecular spectrum can contain a large chemical separation between groups of sites together with a much smaller splitting inside one group.

\begin{figure}[!t]
  \centering
  \includegraphics[width=\columnwidth,height=0.65\textheight,keepaspectratio]{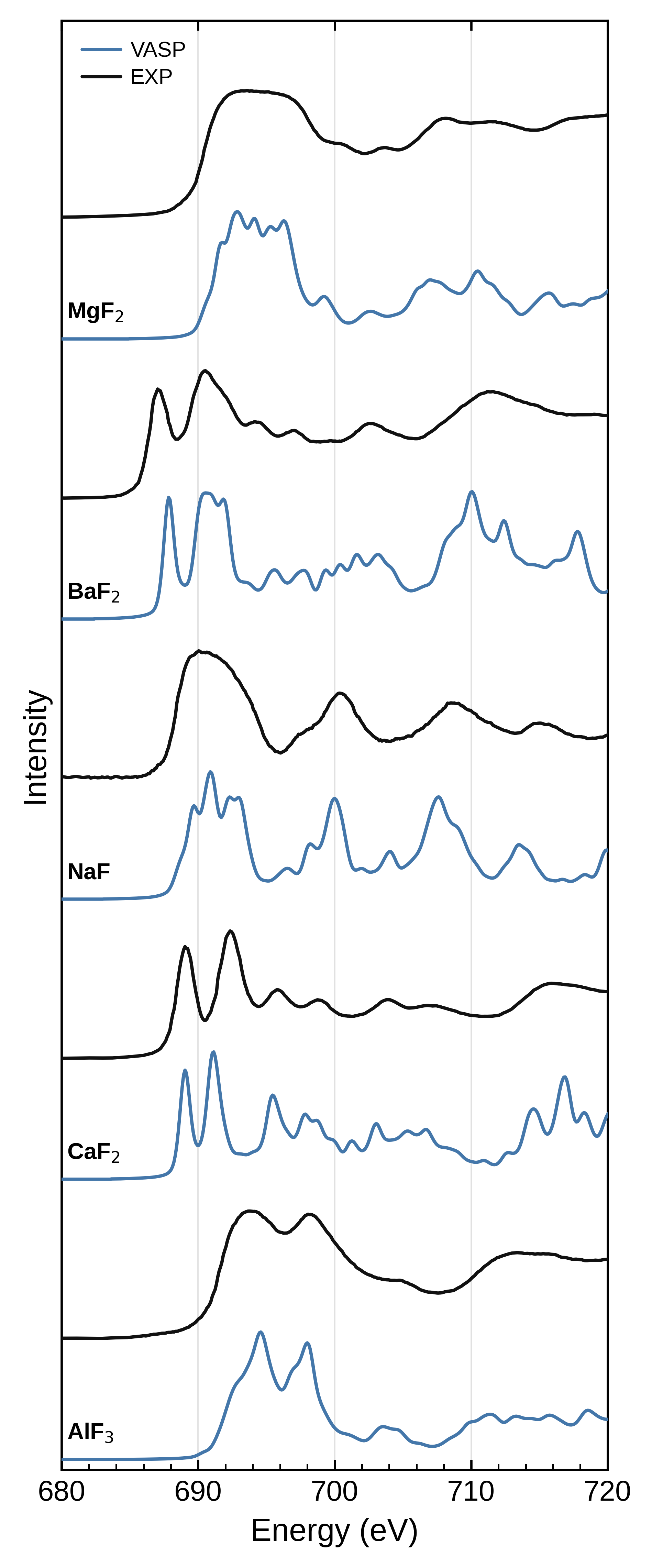}
  \caption{F K-edge XANES of $\alpha$-AlF$_3$, $\alpha$-CaF$_2$, $\alpha$-NaF, $\alpha$-BaF$_2$, and $\alpha$-MgF$_2$, arranged from bottom to top. Experimental spectra \cite{ishii_integration_2023} are black and PAW-XCH spectra are blue. The computed $\alpha$-CaF$_2$ spectrum is aligned to experiment and the same energy shift is applied to the other computed spectra. The experimental alignment is used only for visualization and is separate from the construction of Eq.~\eqref{eq_final_formula}.}
  \label{fig_f_edge_spectra}
\end{figure}

The same relation also compares a fixed edge across materials. Figure~\ref{fig_f_edge_spectra} and Table~\ref{tab_f_shift} examine five fluoride solids spanning covalent and ionic F coordination. Referencing all systems to $\alpha$-CaF$_2$ removes the common F K-edge atomic terms and leaves differences in the screened supercell response.

Table~\ref{tab_f_shift} lists the pairwise $\ddEval$ values referenced to $\alpha$-CaF$_2$. The calculated ordering is $\alpha$-AlF$_3$ ($+1.62\eV$), $\alpha$-MgF$_2$ ($+0.52\eV$), $\alpha$-CaF$_2$ ($0.00\eV$), $\alpha$-BaF$_2$ ($-1.01\eV$), and $\alpha$-NaF ($-1.18\eV$). The series spans $2.80\eV$, with the more covalent $\alpha$-AlF$_3$ at the highest relative edge energy. The same displacement is applied to all spectra after aligning $\alpha$-CaF$_2$ to experiment, so their relative shifts remain fixed by $\ddEval$.

\begin{table}[t]
\caption{F K-edge double difference $\ddEval$ referenced to $\alpha$-CaF$_2$.}
\label{tab_f_shift}
\begin{ruledtabular}
\begin{tabular}{lr}
System & $\ddEval$ relative to $\alpha$-CaF$_2$ (eV) \\
\hline
$\alpha$-AlF$_3$ & $+1.62$ \\
$\alpha$-MgF$_2$ & $+0.52$ \\
$\alpha$-CaF$_2$ & $\phantom{+}0.00$ \\
$\alpha$-BaF$_2$ & $-1.01$ \\
$\alpha$-NaF     & $-1.18$ \\
\end{tabular}
\end{ruledtabular}
\end{table}

\begin{figure}[!t]
  \centering
  \includegraphics[width=\columnwidth,height=0.58\textheight,keepaspectratio]{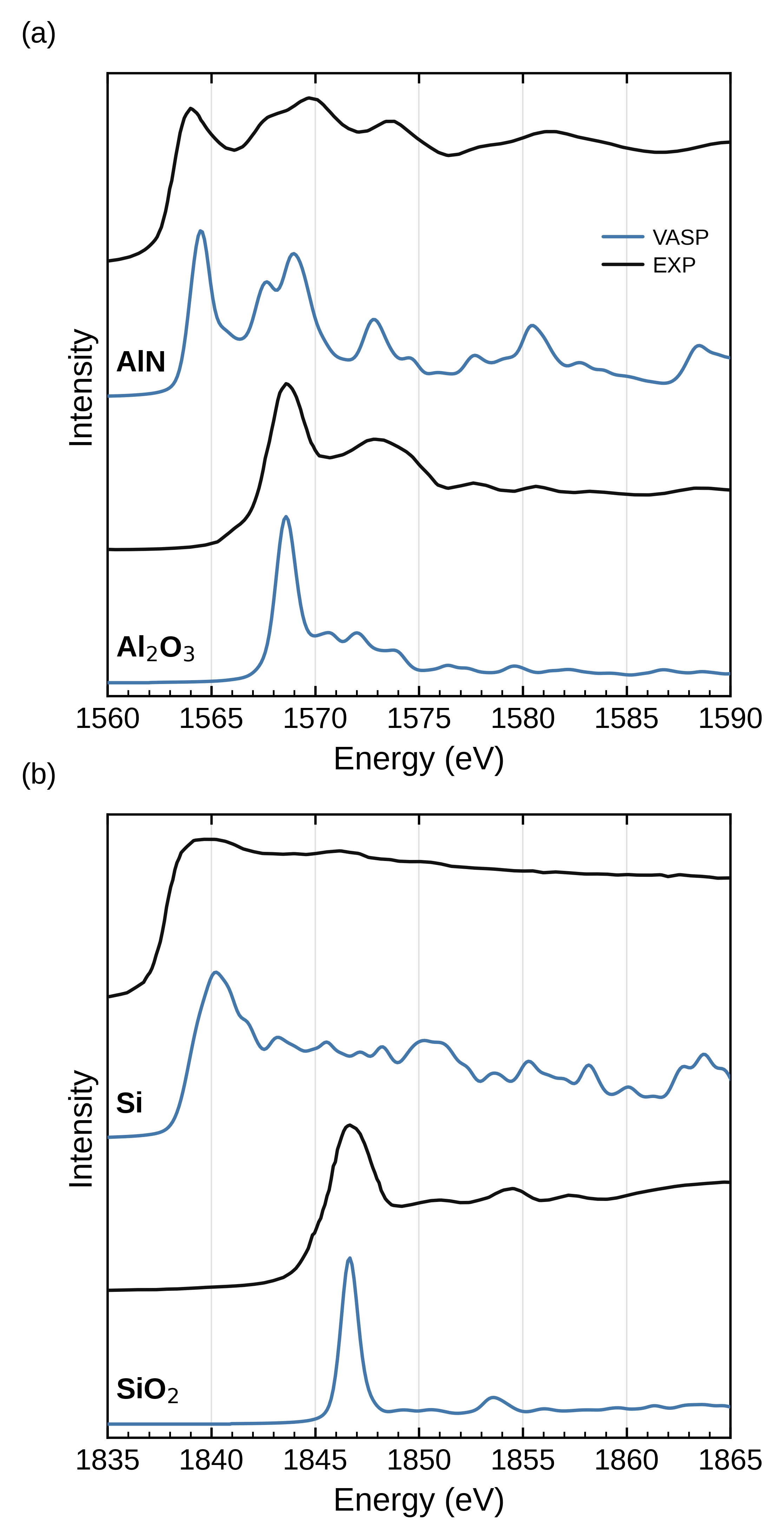}
  \caption{Same-element K-edge comparison. (a)~w-AlN \cite{balasubramanian_characterization_2006} and $\alpha$-Al$_2$O$_3$ \cite{murao_thermodynamic_2018}. (b)~Bulk Si and $\alpha$-SiO$_2$ \cite{wu_facile_2017}. Experimental spectra are black and PAW-XCH spectra are blue. In each panel the computed oxide spectrum is aligned to experiment and the same energy shift is applied to the paired computed spectrum. The experimental alignment is used only for visualization and is separate from the construction of Eq.~\eqref{eq_final_formula}.}
  \label{fig_alsi_k}
\end{figure}

Figure~\ref{fig_alsi_k} compares the Al and Si K edges.

For the Al K edge the supercell total-energy differences are $\dEval = 80.07\eV$ for $\alpha$-Al$_2$O$_3$ and $77.94\eV$ for w-AlN, so $\ddEval = 2.13\eV$ places the oxide edge above the nitride edge. For the Si K edge the values are $\dEval = 68.91\eV$ for $\alpha$-SiO$_2$ and $62.21\eV$ for bulk Si, giving $\ddEval = 6.70\eV$ between the oxide and the elemental semiconductor. The larger Si shift is consistent with the pronounced change from elemental Si to the oxygen-coordinated network. Both relative orderings agree with the experimental trends shown in Fig.~\ref{fig_alsi_k}.

\FloatBarrier
\subsection{PAW-Dataset Dependence at the Al and Si L\texorpdfstring{$_{2,3}$}{2,3} Edges}
\label{sec_l23_results}

\begin{figure*}[!p]
  \centering
  \includegraphics[width=0.86\textwidth]{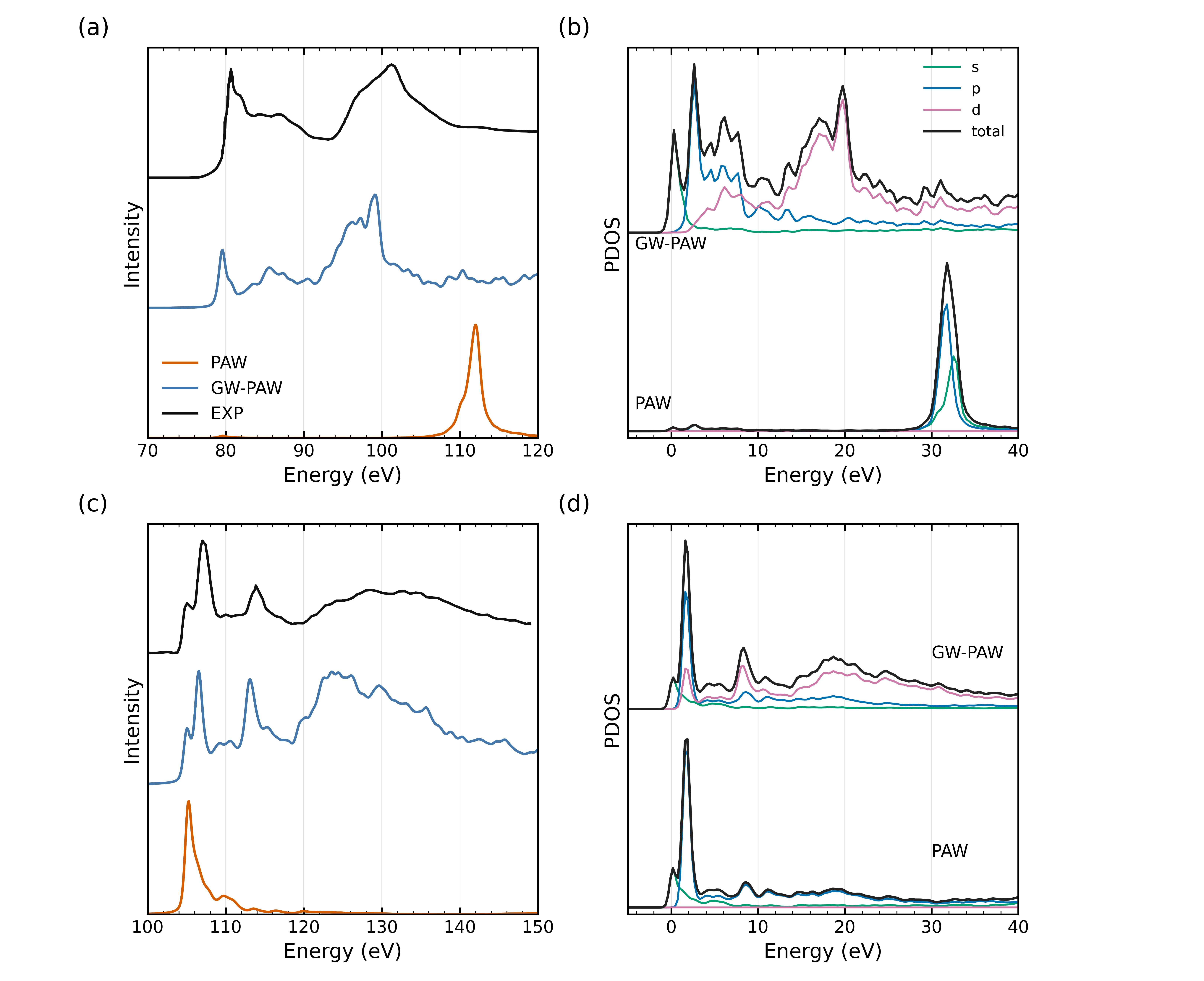}
  \caption{Al and Si L$_{2,3}$ spectra and XCH-state projected densities of states. (a) and (c)~Experimental spectra of $\alpha$-Al$_2$O$_3$ \cite{jiang_situ_2011} and $\alpha$-SiO$_2$ \cite{garvie_use_1994} with XCH spectra from standard PAW and GW-type PAW datasets. The GW-type spectrum is aligned to experiment and the same energy shift is applied to the standard PAW spectrum. (b) and (d)~Absorbing-atom $s$, $p$, and $d$ projections. The energy zero is the XCH Fermi level. Densities from the two datasets share one scale and are vertically offset for clarity.}
  \label{fig_l23_spectra}
\end{figure*}

The merged L$_{2,3}$ construction and its scope are defined in Sec.~\ref{sec_l23_method}. Figure~\ref{fig_l23_spectra} compares PBE-based XCH calculations for the Al and Si L$_{2,3}$ edges using two alternative PAW datasets: the standard PAW dataset and the GW-type PAW dataset. Here, ``GW-type'' denotes only the PAW-dataset construction; no GW calculation was performed, and no GW self-energy correction was applied.

The standard PAW spectra in Fig.~\ref{fig_l23_spectra}(a,c) contain a sharper and simpler near-edge response than experiment. At the Al edge the dominant low-energy feature is concentrated into a narrow peak, and the Si edge shows a similarly reduced multiplet structure. Spectra obtained with the GW-type PAW datasets show a broader multi-peak response with closer main-peak positions, splitting, and intensity distribution. These datasets retain a richer representation of unoccupied partial waves. Because the same energy shift is applied to both computed spectra in each panel, their relative displacement and shape difference remain visible.

The projected densities of states (PDOS) in Fig.~\ref{fig_l23_spectra}(b,d) provide a microscopic comparison. Dipole transitions from a $2p$ core level sample low-lying unoccupied states with substantial $s$ and $d$ character. The standard PAW calculation gives less low-energy $d$-projected weight and concentrates the remaining weight into a narrower interval, consistent with its sharp near-edge response. GW-type PAW gives a larger and broader distribution of this weight, which correlates with the observed multi-peak spectrum. The $s$ and $p$ projections also change, although the most visible dataset contrast occurs in the low-energy $d$ channel. This correlation supports a role for low-lying $3d$-like unoccupied states. A complete assignment would additionally require the energy-dependent transition matrix elements and their coupling to the final-state wave functions.

At the matched K edges, standard and GW-type PAW datasets produce only a small difference in $\dEval$ and no substantial change in the raw relative edge position. This control indicates that the pronounced L$_{2,3}$ response is associated mainly with the PAW description of low-lying unoccupied states and spectral shape. The result also shows why a dataset can improve a near-edge line shape without supplying a separate many-body energy correction. The L$_{2,3}$ dataset sensitivity is distinct from the atomic-reference conclusions for the C, N, O, and F K edges, which use localized $1s$ holes and form the principal absolute-energy validation set.

Within this common XCH framework, the improved multi-peak structure obtained with the GW-type datasets provides evidence for a more suitable unoccupied-state representation. The present PDOS analysis supports this assignment at the level of projected weights and leaves a fully matrix-element-resolved decomposition for future work.

\subsection{Boundary Cases and Scope of the Present Correction}
\label{sec_boundary_cases}
\label{sec_limits}

The all-electron solver includes spectator-core relaxation for the converged K, L$_1$, and L$_{2,3}$ configurations across the second through fourth periods. Three groups nevertheless lie outside the validated set because of limitations in the auxiliary atomic-reference construction. For the Li, Be, and B K edges, the weakly bound $2s$ radial state lies close to the continuum threshold and the present eigenvalue search cannot bracket it stably. For Na L$_1$ and Na L$_{2,3}$, the channel ordering in the semicore-rich \texttt{Na\_sv\_GW} dataset is incompatible with the initialization assumption. The dataset treats $2s$, $2p$, and $3s$ states as semicore or valence channels, while the current initialization assumes a fixed correspondence between occupied PAW channels and atomic-state order. The resulting angular-momentum mapping fails before a valid atomic reference is formed. The periodic XCH and USPP reference calculations remain separate from these auxiliary-solver limitations.

The Al and Si L$_{2,3}$ edges are excluded from the primary absolute-energy validation for the reasons in Secs.~\ref{sec_l23_method} and \ref{sec_l23_results}. Reliable absolute transition energies for these edges require spin-orbit resolution, a non-spherical core-hole potential, a more complete relativistic treatment, and a suitable PAW dataset.

These requirements are coupled. Spin-orbit splitting defines separate initial levels, while the angular character of a $2p$ hole can change the relaxed atomic response. A scalar spherical reference can describe a centroid-like quantity, but it cannot by itself determine two resolved edge positions. The PAW dataset must also represent the low-energy final states that control the observed near-edge structure.

The core-valence decomposition also depends on its reference convention. The all-electron shift $\dEatomall$ depends on the element, prescribed core-hole occupation, atomic occupation convention, and exchange-correlation functional. The residual term $\dEAEref$ additionally depends on the VASP PAW dataset and internal reference convention. These dependencies permit an element-level reference within one consistent scheme. Transfer between datasets or occupation choices remains limited. Element-level reference terms must be recomputed when these settings change, and their numerical values should be reported with the PAW dataset and the core-hole construction.

% ============================================================
\section{Conclusion}
\label{sec_conclusion}
% ============================================================

Fixed-reference PAW-XCH calculations retain one VASP atomic reference when the absorbing-atom core occupation changes. This omits an all-electron atomic reference contribution from the transition-energy axis. We reconstruct that contribution with a self-consistent all-electron single-atom reference and the residual VASP correction $\dEAEref$ of Eq.~\eqref{eq_ae_ref_final}. Combining this term with the direct one-center shift, the supercell total-energy difference, and replacement of the raw single-particle reference yields the absolute axis in Eq.~\eqref{eq_final_formula}.

Single-atom tests spanning K, L$_1$, and L$_{2,3}$ core holes support the numerical consistency of the all-electron atomic reference. For the C, N, O, and F K edges, independent implementations reproduce $\dEatomall$ within about $0.02\eV$. Representative spectra approach the experimental absolute energy scale after the correction is applied. The direct core-only quantity shows the expected reference-convention dependence. For systems sharing the same element, edge, and core-hole scheme, the element-level atomic terms cancel and the chemical shift is given by $\ddEval$.

The Al and Si L$_{2,3}$ spectra are sensitive to the PAW dataset and to low-energy $3d$-like unoccupied states. GW-type PAW produces broader $d$-projected weight and a near-edge multiplet structure closer to experiment, while the present scalar atomic reference does not resolve the spin-orbit-split components. Additional limits arise from radial-solver failures for Li, Be, and B K edges, channel mapping for Na L$_1$ and Na L$_{2,3}$, and reference dependence of the core-only decomposition. Changes in the PAW dataset, occupation scheme, or functional require recomputation of the element-level references. The rigid correction also leaves relative peak-spacing errors unchanged.

% ============================================================

%  Acknowledgments
% ============================================================
\begin{acknowledgments}
This work was partially supported by the Project for a Common Platform to Expand Hydrogen Utilization (P25002, 250105), commissioned by the New Energy and Industrial Technology Development Organization (NEDO).
Y. Wang acknowledges support from the Program for Leading Graduate Schools (MERIT-WINGS).
\end{acknowledgments}

% ============================================================
%  Appendices
% ============================================================
\FloatBarrier
\appendix

% ============================================================
\section{Glossary of Physical Quantities}
\label{app_glossary}

Table~\ref{tab_glossary} lists the notation and its primary definition.

\begin{table*}[t]
\caption{Glossary of the physical quantities used in the manuscript, grouped by the total-energy decomposition to which they belong.
All shifts use the XCH-minus-GS sign convention of Eq.~\eqref{eq_sign_convention} unless noted otherwise.}
\label{tab_glossary}
\begin{ruledtabular}
\begin{tabular}{lll}
Symbol & Physical meaning & Defined in \\
\hline
$\ETE$ & Transition energy at the absorption edge & Eq.~\eqref{eq_energy_conservation} \\
$\dEcoreatom$ & Atomic core-reference contribution to $\ETE$ & Eq.~\eqref{eq_energy_conservation} \\
$\dEval$ & Supercell total-energy difference (XCH$-$GS) for the site & Sec.~\ref{sec_decomp} \\
$\ddEval$ & Double difference of $\dEval$, the same-element chemical shift & Sec.~\ref{sec_chemical_shift} \\
$\wraw$ & Raw XCH spectral axis, positioned by the internal transition & Sec.~\ref{sec_final_formula} \\
$\wcorr$ & Corrected transition-energy axis & Eq.~\eqref{eq_final_formula} \\
$\varepsilon_{\mathrm{lumo}}^{\mathrm{XCH}}$ & Kohn-Sham energy of the lowest unoccupied state in XCH & Sec.~\ref{sec_final_formula} \\
$\varepsilon_{\mathrm{core}}^{\mathrm{GS}}$ & Kohn-Sham energy of the absorbing core level in GS & Sec.~\ref{sec_final_formula} \\
\hline
$\dEatomall$ & All-electron single-atom total-energy shift & Sec.~\ref{sec_atom_core} \\
$\dEatomval$ & Pseudo-valence single-atom reference shift & Sec.~\ref{sec_atom_core} \\
$\dEuspp$ & USPP core-valence partition (all-electron minus pseudo-valence) & Eq.~\eqref{eq_uspp_partition} \\
$\dEpaw$ & Direct VASP one-center core-reference shift & Sec.~\ref{sec_atom_core} \\
$\dEisoVASP$ & Isolated-atom VASP total-energy shift, fixed vacuum reference & Eq.~\eqref{eq_iso_atom_def} \\
$\dEvalatomVASP$ & VASP single-atom valence reference shift entering $\dEAEref$ & Sec.~\ref{sec_atom_core} \\
$\dEAEref$ & Residual atomic reference correction on the corrected axis & Eq.~\eqref{eq_ae_ref_final} \\
$Z_{\mathrm{core}}$ & Core electrons below the active hole, sets the edge branch & Sec.~\ref{sec_branch_logic} \\
\end{tabular}
\end{ruledtabular}
\end{table*}

% ============================================================
\section{Radial Construction of the Atomic Reference Energies}
\label{app_S1}

This appendix gives the radial construction of the VASP one-center core reference and the all-electron single-atom reference introduced in Sec.~\ref{sec_atom_core}. It records the density, energy-component, convergence, and integration details required for reproduction.

\subsection{VASP One-Center Core Reference}
\label{app_S1_paw_transform}

In the PAW formalism, the all-electron wave function is recovered from the pseudo-wave function through a linear transformation \cite{blochl_projector_1994,kresse_ultrasoft_1999},
\begin{equation}
  |\psi_n\rangle
  = |\tilde\psi_n\rangle
  + \sum_{a,i}\left(|\phi_i^a\rangle-|\tilde\phi_i^a\rangle\right)
    \langle \tilde p_i^a | \tilde\psi_n \rangle .
  \label{eq_paw_transform}
\end{equation}
The corresponding one-center density matrix is
\begin{equation}
  \rho_{ij}^{a}=\sum_n f_n
  \langle \tilde\psi_n|\tilde p_i^a\rangle
  \langle \tilde p_j^a|\tilde\psi_n\rangle ,
  \label{eq_paw_onecenter}
\end{equation}
which determines the all-electron-minus-pseudo-density difference within the augmentation sphere.
The PAW one-center core reference needs only the spherical one-center core-electron energy term at the absorbing atom, so the valence reference density may be held fixed at the spherically averaged background given by the PAW atomic data.
The site index $a$ selects the absorbing atom, while $i$ and $j$ label its partial waves and projectors. The transformation supplies the local all-electron information associated with the stored dataset. For the present atomic reference, the nonspherical response of the periodic valence density is excluded from the radial problem. The fixed spherical background instead reproduces the element and dataset convention used to define the one-center core energy.
In the $k$-th core-occupation state, the total radial reference density is written as
\begin{equation}
  \begin{split}
  n_{\mathrm{atom}}^{(k)}(r)
  &= n_{\mathrm{core,atom}}^{(k)}(r)
  + n_{\mathrm{val,atom}}^{\mathrm{PAW,ref}}(r), \\
  n_{\mathrm{core,atom}}^{(k)}(r)
  &= \frac{1}{4\pi r^2}\sum_{i\in\mathrm{core}} f_{i}^{(k)}
  \left|u_{i}^{(k)}(r)\right|^2 ,
  \end{split}
  \label{eq_radial_density}
\end{equation}
where $n_{\mathrm{val,atom}}^{\mathrm{PAW,ref}}(r)$ is the spherically averaged PAW valence reference density and $f_{i}^{(k)}$ is the occupation of core shell $i$ in state $k$.
The sum over $i\in\mathrm{core}$ runs over all core shells present in the PAW dataset, not only the active hole shell.
For the light-element K edges with $Z_{\mathrm{core}}\le 2$ it usually contains only the $1s$ shell, whereas for $Z_{\mathrm{core}}>2$ it also contains the spectator core shells.
Within this fixed valence background, the radial Kohn-Sham equation for the core shells is solved \cite{kohn_self-consistent_1965},
\begin{equation}
  \left[-\frac{1}{2}\frac{d^2}{dr^2}
  +\frac{l(l+1)}{2r^2}+V_{\mathrm{eff,atom}}^{(k)}(r)\right]
  u_{i}^{(k)}(r)
  = \epsilon_{i}^{(k)}u_{i}^{(k)}(r),
  \label{eq_radial_ks}
\end{equation}
where the effective potential $V_{\mathrm{eff,atom}}^{(k)}$ is generated from the total spherical density $n_{\mathrm{core,atom}}^{(k)}+n_{\mathrm{val,atom}}^{\mathrm{PAW,ref}}$ of Eq.~\eqref{eq_radial_density}.
The valence reference density $n_{\mathrm{val,atom}}^{\mathrm{PAW,ref}}$ is a fixed input and is not reoptimized during the self-consistency loop.
Only the core contribution to the total spherical density is updated in this potential and mixed linearly,
\begin{equation}
  n_{\mathrm{core,atom}}^{(m+1)}(r)
  = (1-\alpha)\,n_{\mathrm{core,atom}}^{(m)}(r)
  + \alpha\, n_{\mathrm{core,atom}}^{\mathrm{new}}(r) ,
  \label{eq_radial_mixing}
\end{equation}
and the convergence criterion is the maximum core-density difference on the radial grid,
\begin{equation}
  R^{(m)}
  = \max_r
  \left| n_{\mathrm{core,atom}}^{\mathrm{new}}(r)
  - n_{\mathrm{core,atom}}^{(m)}(r) \right| .
  \label{eq_radial_residual}
\end{equation}

The same radial grid, mixing rule, and convergence criterion are used for the GS and core-hole occupations. Each iteration rebuilds the effective potential from the updated core density and the fixed PAW valence reference. All core shells in the dataset are then solved in that potential. This procedure allows spectator core shells to respond to the active hole while leaving the pseudo-valence reference unchanged.

The VASP one-center core-reference energy in state $k$ is
\begin{equation}
  \begin{split}
  E_{\mathrm{core,atom}}^{\mathrm{VASP},(k)}
  &= \sum_{i\in\mathrm{core}} f_i^{(k)}\epsilon_i^{(k)}
  - \int n_{\mathrm{core,atom}}^{(k)}
  V_{\mathrm{eff,atom}}^{(k)}\,d^3r \\
  &\quad + E_{\mathrm{H}}\!\left[n_{\mathrm{core,atom}}^{(k)}\right]
  + E_{\mathrm{en}}\!\left[n_{\mathrm{core,atom}}^{(k)}\right] \\
  &\quad + E_{\mathrm{xc}}\!\left[n_{\mathrm{core,atom}}^{(k)}\right]
  - E_{0,Z}^{\mathrm{PAW}} .
  \end{split}
  \label{eq_epaw_def}
\end{equation}
The quantity $E_{0,Z}^{\mathrm{PAW}}$ is the fixed element reference constant stored in the PAW dataset. It cancels between GS and XCH evaluations that use the same dataset. Equation~\eqref{eq_epaw_def} defines the VASP one-center core-reference functional. The fixed valence reference density contributes through the effective potential, core eigenvalues, and kinetic term, while the Hartree, electron-nucleus, and exchange-correlation terms use the core density. This functional supplies the reference change missing from the fixed-dataset calculation and remains distinct from a complete free-atom all-electron total energy.
The eigenvalue sum and potential subtraction form the kinetic contribution after the effective-potential expectation value has been removed. $E_{\mathrm{H}}$ is the core-density Hartree energy, $E_{\mathrm{en}}$ is its interaction with the point nucleus of charge $Z$, and $E_{\mathrm{xc}}$ is evaluated from the core density with the chosen functional. The fixed valence density affects these core orbitals through $V_{\mathrm{eff,atom}}^{(k)}$ and is not inserted as an additional density argument in the last three terms.
The four explicit energy components are
\begin{align}
  E_{\mathrm{kin}}^{\mathrm{core}}
  &= \sum_{i\in\mathrm{core}} f_i^{(k)}\epsilon_i^{(k)}
  - \int n_{\mathrm{core,atom}}^{(k)}(\mathbf r)V_{\mathrm{eff,atom}}^{(k)}(\mathbf r)d^3r,
  \label{eq_ekin_core}\\
  E_{\mathrm{H}}^{\mathrm{core}}
  &= \frac{1}{2}\iint
  \frac{n_{\mathrm{core,atom}}^{(k)}(\mathbf r)n_{\mathrm{core,atom}}^{(k)}(\mathbf r')}{|\mathbf r-\mathbf r'|}
  d^3r d^3r',
  \label{eq_eh_core}\\
  E_{\mathrm{en}}^{\mathrm{core}}
  &= -Z\int\frac{n_{\mathrm{core,atom}}^{(k)}(\mathbf r)}{r}d^3r,
  \label{eq_een_core}\\
  E_{\mathrm{xc}}^{\mathrm{core}}
  &= \int n_{\mathrm{core,atom}}^{(k)}
  \varepsilon_{\mathrm{xc}}\!\left[n_{\mathrm{core,atom}}^{(k)},|\nabla n_{\mathrm{core,atom}}^{(k)}|\right]d^3r .
  \label{eq_exc_core}
\end{align}
The components are evaluated over the range in Sec.~\ref{app_S1_radial}. The GS-to-XCH difference of Eq.~\eqref{eq_epaw_def} is the direct shift $\dEpaw$. For the C, N, O, and F K edges it is the sum of the component changes under $1s^2\to1s^1$,
\begin{equation}
  \dEpaw
  = \sum_{\alpha\in\{\mathrm{kin,H,en,xc}\}}
  \left(E_{\alpha}^{\mathrm{core}}\big|_{1s^1}
  -E_{\alpha}^{\mathrm{core}}\big|_{1s^2}\right).
  \label{eq_shift_components}
\end{equation}

\subsection{All-Electron Single-Atom Reference}
\label{app_S1_ae_ref}

The scalar-relativistic spherical all-electron solver evaluates $\dEatomall$ separately from the one-center functional. It uses the actual nuclear charge and the GS or core-hole occupations defined in Sec.~\ref{sec_atom_core}. All occupied shells enter the self-consistent density
\begin{equation}
  n_{\mathrm{all,atom}}^{\mathrm{AE},(k)}(r)
  = \frac{1}{4\pi r^2}\sum_i f_i^{(k)}\left|u_i^{(k)}(r)\right|^2 ,
  \label{eq_ae_density_app}
\end{equation}
where the active core shell, spectator core shells, and occupied valence shells are solved in one potential. For $Z_{\mathrm{core}}>2$, valence and core shells relax together in response to the core hole.
The radial orbitals are normalized on the all-electron grid and weighted by their prescribed occupations $f_i^{(k)}$. The potential is rebuilt from the complete density in Eq.~\eqref{eq_ae_density_app}, so the valence density is a variational part of this calculation. This differs from the fixed valence background used by the one-center construction in Sec.~\ref{app_S1_paw_transform}. The use of one spherical potential keeps the active core, spectator core, and occupied valence responses mutually self-consistent.
The all-electron total energy for each state is written as the eigenvalue sum plus the double-counting correction,
\begin{equation}
  E_{\mathrm{all,atom}}^{\mathrm{AE},(k)}
  = \sum_i f_i^{(k)}\epsilon_i^{(k)}
  + E_{\mathrm{dc}}\!\left[n_{\mathrm{all,atom}}^{\mathrm{AE},(k)}\right] ,
  \label{eq_eatom_all_app}
\end{equation}
where $E_{\mathrm{dc}}$ contains the Hartree, electron-nucleus, and exchange-correlation double-counting terms. The nuclear charge is passed independently of the occupation-derived electron number, which keeps the core-hole atom at the elemental value $Z$.
The total-energy difference is formed only after both occupations have converged. Passing $Z$ independently is required because the core-hole configuration is ionized and contains fewer electrons than the GS atom. The separation preserves the electron-nucleus interaction of the original element while allowing the missing core electron to change the self-consistent density and eigenvalue sum.

\label{app_S1_iso}
The isolated-atom quantities in Eq.~\eqref{eq_iso_atom_def} use one atom in a vacuum supercell large enough to suppress interactions with periodic images. The GS calculation has the neutral occupation and the core-hole calculation has the FCH occupation without a compensating valence electron. The single-atom pseudopotential and all-electron calculations then contain the same number of valence electrons. Both states use the exchange-correlation functional, plane-wave cutoff convention, and VASP PAW dataset of the corresponding periodic calculation. The same FCH construction supplies $\dEvalatomVASP$ in Eq.~\eqref{eq_ae_ref_final}.
The vacuum cell is enlarged until the total-energy difference is insensitive to further increases, which keeps the periodic-image contribution below the accuracy required for the reference term. The GS and FCH calculations use the same cell, sampling, and numerical settings. Their subtraction isolates the occupation response under a fixed vacuum reference. This isolated construction is used only for the atomic residual and remains distinct from the charge-neutral XCH supercell that supplies $\dEval$ for a material.

\subsection{Branch Logic for the Two Edge Classes}
\label{app_S1_branch}

The two solver branches are defined in Sec.~\ref{sec_branch_logic}. Their radial quantities are supplied by Secs.~\ref{app_S1_paw_transform} and \ref{app_S1_ae_ref} and enter the common residual correction in Eq.~\eqref{eq_ae_ref_final}.

\subsection{Radial Integration Range}
\label{app_S1_radial}

The energy components of $E_{\mathrm{core,atom}}^{\mathrm{VASP}}$ are integrated over the full radial grid of the atom,
$\{r_1, \ldots, r_{N_{\max}}\}$, with the upper bound $R_{\max}$ taken as the outermost point of the radial grid, and no truncation is applied at the augmentation radius $r_{\mathrm{aug}}$.
For the representative C, N, and O K edges tabulated below, the $1s$ core orbital is highly localized, and its 99\%-charge radius $r_{99\%}$ is much smaller than $r_{\mathrm{aug}}$.

\begin{center}
\begin{tabular}{lrrr}
\toprule
Element & $R_{\max}$ (\AA) & $r_{\mathrm{aug}}$ (\AA) & $r_{99\%}^{1s}\,\mathrm{(GS)}$ (\AA) \\
\midrule
C & 6.16 & 0.70 & 0.41 \\
N & 5.32 & 0.59 & 0.35 \\
O & 4.92 & 0.72 & 0.30 \\
\bottomrule
\end{tabular}
\end{center}

The integration runs to $R_{\max}$ in all cases.
Because $r_{99\%} \ll r_{\mathrm{aug}}$, the $1s$ core density is essentially complete well inside $r_{\mathrm{aug}}$, so the region beyond $r_{\mathrm{aug}}$ carries negligible core-orbital weight and the choice of upper bound does not affect the core-energy components.
The augmentation charge affects only the overall offset of the reference potential through the average-potential alignment, and does not change the integration bounds.

\subsection{Evaluation of the Energy Components}

The kinetic term follows from the core eigenvalues and the effective-potential expectation value. The Hartree and electron-nucleus terms use the radial core density and point charge $Z$, while the PBE exchange-correlation term uses the core density alone. The fixed PAW valence reference affects the eigenvalues and kinetic term through the effective potential. Differences between $\dEpaw$ and $\dEuspp$ consequently combine atomic reference conventions with exchange-correlation partitioning, as discussed in Sec.~\ref{sec_decore_discussion}.
All radial integrals use the quadrature associated with the stored PAW grid. The kinetic expression is evaluated consistently with the eigenproblem in Eq.~\eqref{eq_radial_ks}, and the Coulomb terms use the same point-nucleus convention in both occupations. Evaluating the component changes separately provides a numerical check that their sum reproduces the direct shift $\dEpaw$. It also identifies which terms respond most strongly to a changed core occupation without assigning them an independent observable meaning.

% ============================================================
\section{Computational Details and Datasets}
\label{app_details}
\label{sec_details}
\label{app_S3}

\subsection{Computational Settings}

\paragraph{Organic molecules.}
For all organic molecules, after structural modeling the molecule was placed in a vacuum supercell of
$15\,\text{\AA} \times 15\,\text{\AA} \times 15\,\text{\AA}$
to sufficiently suppress interactions between periodic images.
Structural relaxation was performed with VASP using the PBE exchange-correlation functional, together with the DFT-D3 dispersion correction (Grimme, zero-damping form) \cite{grimme_consistent_2010}.
XCH calculations of the XANES/ELNES absorption spectra were carried out with both VASP and the ultrasoft pseudopotential method, the latter performed with CASTEP \cite{clark_first_2005}, to enable a direct comparison between the two implementations.
All organic-molecule calculations used single-$k$-point ($\Gamma$-point) sampling, with a plane-wave cutoff energy of $500\eV$.
The relaxed geometry was used unchanged for the VASP and CASTEP core-hole calculations so that their atomic reference quantities could be compared without a structural contribution. Separate XCH calculations were performed for each inequivalent absorbing site included in a site-resolved spectrum.

\paragraph{Solids.}
For solid systems, structural relaxation used the PBEsol functional \cite{perdew_restoring_2008}, after which VASP and ultrasoft-pseudopotential XCH calculations were performed on the relaxed structure.
The $k$-point sampling density was $0.03\,\text{\AA}^{-1}$, with a plane-wave cutoff energy of $500\eV$.
The same relaxed cell and atomic positions were used for the paired reference calculations. Site-specific supercells differed only through the selected absorbing atom and its prescribed core-hole occupation.

\paragraph{PAW datasets.}
Production spectra and total-energy quantities used GW-type PAW datasets unless stated otherwise. Standard PAW datasets were used only for Fig.~\ref{fig_l23_spectra} and Appendix~\ref{app_S9}.

\paragraph{Spectral broadening.}
The raw computed XANES spectra were broadened with a pseudo-Voigt function combining a Lorentzian component, which models the core-hole lifetime broadening, and a Gaussian component, which models the instrumental-resolution broadening.
Both components used a full width at half maximum of $0.80\eV$, and the mixing parameter was $\eta = 0.50$, corresponding to an equal Lorentzian and Gaussian weighting.

\section{PAW-Dataset Dependence of the One-Center Core-Reference Shift at the Si K and L\texorpdfstring{$_{2,3}$}{2,3} Edges}
\label{app_S9}

Table~\ref{tab_si_test} shows the PAW-dataset dependence of the direct one-center shift $\dEpaw$, while the Si all-electron shift remains stable within about $0.03\eV$. Standard and GW-type PAW datasets lie within a few eV of the USPP reference. The semicore Si$_{\rm sv,GW}$ dataset moves the K-edge value by more than $550\eV$ because its $2s$ and $2p$ shells enter the pseudo-valence reference. The L$_2$ entries represent the $2p$ core-hole shell. Differences are measured from the USPP values $1902.31\eV$ at K and $149.97\eV$ at L$_{2,3}$.

The standard Si and Si$_{\rm GW}$ datasets keep the relevant inner shells in the core reference and give K-edge shifts of $1903.33$ and $1904.06\eV$. Their L$_{2,3}$ values are $151.57$ and $152.45\eV$. In the Si$_{\rm sv,GW}$ dataset, moving the $2s$ and $2p$ shells into the valence reference changes the bookkeeping of the one-center core quantity and produces $2455.81\eV$ at K. The all-electron atomic calculation uses the actual Si nuclear charge and occupation independently of that PAW partition, which explains its stability across the dataset comparison.

This comparison illustrates why the PAW dataset must accompany any tabulated value of $\dEpaw$. A change in the core-valence boundary modifies the one-center reference even when the chemical element and nominal absorption edge remain fixed. The corresponding residual correction must be recomputed under the same dataset convention before it is used on an absolute axis.

\begin{table}[b]
\caption{Direct one-center shift $\dEpaw$ for Si PAW datasets at the K and L$_{2,3}$ edges. Differences are relative to the USPP reference $\dEuspp$.}
\label{tab_si_test}
\begin{ruledtabular}
\begin{tabular}{lcrr}
PAW dataset & Edge &
$\dEpaw$ (eV) &
Difference (eV) \\
\hline
Si                 & K          & 1903.33 & $+1.02$ \\
Si$_{\rm GW}$      & K          & 1904.06 & $+1.75$ \\
Si$_{\rm sv,GW}$   & K          & 2455.81 & $+553.50$ \\
Si                 & L$_{2,3}$  & 151.57 & $+1.60$ \\
Si$_{\rm GW}$      & L$_{2,3}$  & 152.45 & $+2.48$ \\
\end{tabular}
\end{ruledtabular}
\end{table}

\FloatBarrier
% ============================================================
%  References
% ============================================================
\revtexbibstylebeforebibdata
\bibliography{references}

\end{document}